\documentclass[preprint,aps,nofootinbib,superscriptaddress,showpacs]{revtex4}
\usepackage{graphicx}
\usepackage{epsfig}
\usepackage{bm}
\usepackage{amssymb}

\newcommand{\be}{\begin{eqnarray}}
\newcommand{\ee}{\end{eqnarray}}

\begin{document}

\preprint{ADP-12-49/T816}

\title{Asymmetry in the neutrino and anti-neutrino reactions in a nuclear medium 
}

\author{Myung-Ki Cheoun}
 \email{cheoun@ssu.ac.kr}
\author{Kiseok Choi}
\affiliation{Department of Physics, Soongsil University, Seoul 156-743, Korea
}
\author{K. S. Kim}%
\affiliation{School of Liberal Arts and Science, Korea Aerospace University, Koyang 412-791, Korea
}%
\author{Koichi Saito}
\affiliation{Department of Physics, Faculty of Science and Technology, Tokyo University of Science, Noda 278-8510, Japan
}%
\author{Toshitaka Kajino}
\affiliation{National Astronomical Observatory of Japan, 2-21-1 Osawa, Mitaka, Tokyo 181-8588, Japan
}%
\author{Kazuo Tsushima}
\affiliation{CSSM, School of Chemistry and Physics, University of Adelaide, Adelaide SA 5005, Australia
}%
\author{Tomoyuki Maruyama}
\affiliation{College of Bioresource Sciences, Nihon University, Fujisawa 252-8510, Japan
}%

\date{\today}
\begin{abstract}
We study the effect of the density-dependent axial and vector form factors
on the electro-neutrino ($\nu_e$) and anti-neutrino $({\bar \nu}_e)$ reactions
for a nucleon in nuclear matter or in $^{12}$C.
The nucleon form factors in free space are presumed to be modified for a bound nucleon in a nuclear medium.
We adopt the density-dependent form factors calculated by the quark-meson coupling (QMC) model,
and apply them to the $\nu_e$ and ${\bar \nu}_e$ induced reactions with the initial energy $E = $ 8 $\sim$ 80 MeV.   We find that
the total ${\nu}_e$ cross sections on $^{12}$C as well as a nucleon in nuclear matter are reduced by
about 5 \% at the nuclear saturation density, $\rho_0$.
This reduction is caused by the modification of the nucleon structure in matter.
Although the density effect for both cases is relatively small, it is comparable with the effect of Coulomb distortion
on the outgoing lepton in the $\nu$-reaction.
In contrast, the density effect on the ${\bar \nu}_e$ reaction reduces
the cross section significantly in both nuclear matter and $^{12}$C cases, and the amount maximally becomes of about 35 \% around $\rho_0$.
Such large asymmetry in the $\nu_e$ and ${\bar \nu}_e$ cross sections, which seems to be nearly independent of the target,
is originated from the difference in the helicities of ${\bar \nu}_e$ and ${\nu}_e$.
It is expected that the asymmetry influences the r-process and also the neutrino-process nucleosynthesis in core-collapse supernovae.
\end{abstract}

\keywords{neutrino-induced reactions, weak and electro-magnetic form factors, nuclear matter, proto-neutron stars, density dependence}
\pacs{14.20.Dh, 25.30.Pt, 26.30.-k, 26.30.Jk}
\maketitle

In the core-collapse supernova explosion, the
neutrino ($\nu$) heating was suggested as one of the main mechanisms for the explosion, leading to the
so-called $\nu$ driven explosions. The central object formed at the core bounce is expected to be a hot
and lepton-rich proto-neutron star (PNS).
Therefore, the shock propagation by the initial bounce from
the contraction of heavy nuclei crosses the $\nu$-sphere, {\it i.e.} the neutrino energy- and flavor-dependent
sphere, and releases vast numbers of neutrinos~\cite{Fisc11}.

These neutrinos propagate through the PNS, whose density is believed to be of about a few times of the nuclear saturation density, $\rho_0 \sim 0.15 {fm}^{-3} $.
During the propagation, the neutrinos interact with nucleons in a dense nuclear medium.
For example, the asymmetry in $\nu$ scattering and absorption in a magnetized PNS may account for the pulsar kick of neutron stars, according to
the detailed study of the $\nu$ transport in dense matter by a relativistic mean field theory~\cite{Maru11,Maru12}.

Outside of the PNS, the emitted neutrinos also interact with nucleons and the nuclei already produced by the s-process in the progenitor and/or the r-process in the explosion.
Around the Si layer, the anti-neutrino (${\bar \nu}$) absorption in the proton-rich environment may produce neutrons immediately captured by the neutron-deficient nuclei,
which affect the proton process, dubbed as the $\nu p$ process~\cite{Wana11}.
In the O-Ne-Mg layer, the $\nu$ induced reactions may play an important role for producing some p-nuclei which are odd-odd neutron-deficient nuclei.
For example, 
$^{180}$Ta and $^{138}$La in the cosmos are believed to be produced from the $\nu$ process~\cite{Ch10-2,Ch12}.
Other light nuclei abundances are also closely associated with the neutrino interactions in the He-C layer~\cite{Yosh08}.
Of course, the density of the medium outside the PNS is not so high
compared to that of the PNS. However, the nucleons interacting with the neutrinos are strongly bound in a nucleus,
and the interactions should thus be different from those in free space.

Recently, strong evidence for the modification
of the nucleon structure in nuclear matter has been
reported from the proton electromagnetic form factors
measured in the polarized $({\vec e}, e' {\vec p})$ scattering off $^{16}$O \cite{Malo00} and
$^{4}$He~\cite{Diet01,Stra03,Paol10,Mala11} at MAMI and Jefferson Lab, and also from the study of the neutron
properties in a nuclear medium through the polarized $({\vec e}, e' {\vec n})$ scattering off $^{4}$He~\cite{Cloe09}.
Therefore, it is quite interesting to investigate the possible change in the $\nu$ and ${\bar \nu}$ induced reactions due to
the variation of the nucleon properties, in order to pin down the ambiguity inherent in the nucleon and/or
nuclear structure in the interpretation of various $\nu$ reactions in the cosmos.

In this paper, we adopt the density-dependent weak form factors calculated in
the quark-meson coupling (QMC) model~\cite{QMCboundff,QMChe3ff,QMCmatterff,QMCmatterga}.  The quark mass in a hadron can be related to
the quark condensate $\langle {\bar q} q \rangle$ in vacuum.  The mass (or $\langle {\bar q} q \rangle$) in nuclear matter may then be reduced from
the value in vacuum because of the condensed scalar ($\sigma$) field depending on the nuclear density $\rho$, namely the Lorentz-scalar, attractive interaction in matter.
The decrease of the quark mass in matter
leads to the variation of the baryon internal structure at the quark level.  Such an effect is considered self-consistently in the QMC model.
The model has been successfully applied in studying the properties of hadrons in a nuclear medium,
finite nuclei~\cite{QMCfinite} and hypernuclei~\cite{QMChyp}.
(For a review, see Ref.~\cite{QMCreview}.)

An initial study for the effect of the density-dependent weak-current form factors in nuclear matter
can be found in Ref.~\cite{Kim04}.  However, in that paper, the simple, relativistic Fermi gas model has been used to study the
$^{12}$C$(\nu_{\mu}, \mu^-)$X reaction, for which only the $\nu$-flux averaged data have been measured.

In the present study, the nuclear structure of $^{12}$C and related $\nu$ reactions are treated in terms of
quasi-particle random phase approximation (QRPA)~\cite{ch10,ch10-2}, which enables us to carry out more realistic calculations.
We then compare the result with the observed, energy-dependent cross section for the $^{12}$C$(\nu_{e}, e^-)^{12}$N$_{g.s.}$ reaction.
For more thorough understanding of the density effect, the $\nu_e$ and ${\bar \nu}_e$ reactions on a nucleon
in nuclear matter are also examined in detail, as well as the ${\bar \nu}_e$ reaction on $^{12}$C, because the elementary
process may be convenient and useful to see how the variation of the form factors affects the ${\nu}_e$ and ${\bar \nu}_e$ reactions in matter.


The weak current operator $W^{\mu}$ for the $\nu$ induced reaction takes a form of
$V^{\mu} - A^{\mu}$ in the standard electro-weak
theory. For a free nucleon, the weak current operator respectively comprises the vector, the
axial vector and the pseudo scalar form factors, $F_{i=1, 2}^V (Q^2)$,
$F_A (Q^2)$ and $F_P (Q^2)$:
\begin{equation}
{W}^{\mu}=F_{1}^V (Q^2){\gamma}^{\mu}+ F_{2}^V (Q^2){\frac {i}
{2M}}{\sigma}^{\mu\nu}q_{\nu}  + F_A(Q^2) \gamma^{\mu} \gamma^5 +
{ F_P(Q^2) \over {2M}} q^{\mu} \gamma^5~,
\end{equation}
where $M$ is the nucleon mass in free space, and $Q^2 (= - q^2)$ is the four-momentum transfer squared.
Because of the conservation of the vector current (CVC) and nonexistence of the second class
current, we here ignore the scalar form factor in the vectorial part and the axial-tensor form factor in the axial part.
By the CVC hypothesis, the vector and axial form factors for the charged current ($CC$) are respectively expressed as
\cite{giusti1,musolf}
\begin{equation}
F_{i}^{V, CC } (Q^2) =  F_i^{p} ( Q^2) -
 F_i^{n}( Q^2)~,  \ \ \ \ \ F_A^{CC} (Q^2)  = - g_A / {( 1 + Q^2
/ M_A^2)}^2~,  \label{gs}
\end{equation}
where $F_i^{p (n)}$ is the proton (neutron) form factor, and $g_A$ and $M_A$ are the axial coupling constant and the
axial cut off mass, respectively.

Before applying to the $\nu$ reaction, we briefly discuss the change of the in-medium nucleon properties
calculated by the QMC model~\cite{QMCboundff,QMChe3ff,QMCmatterff,QMCmatterga}.
%
%
\begin{figure}
\centering
\includegraphics[width=8.0cm]{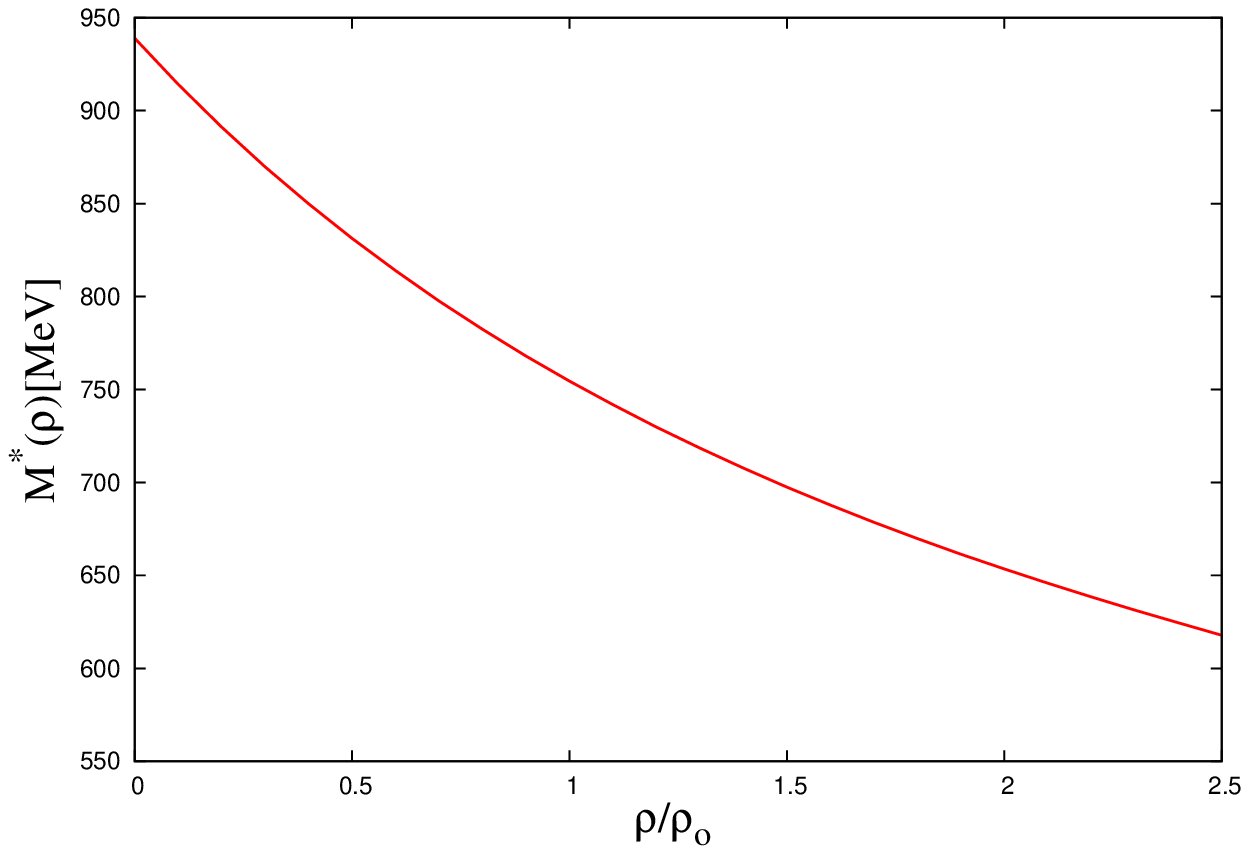}
\includegraphics[width=8.0cm]{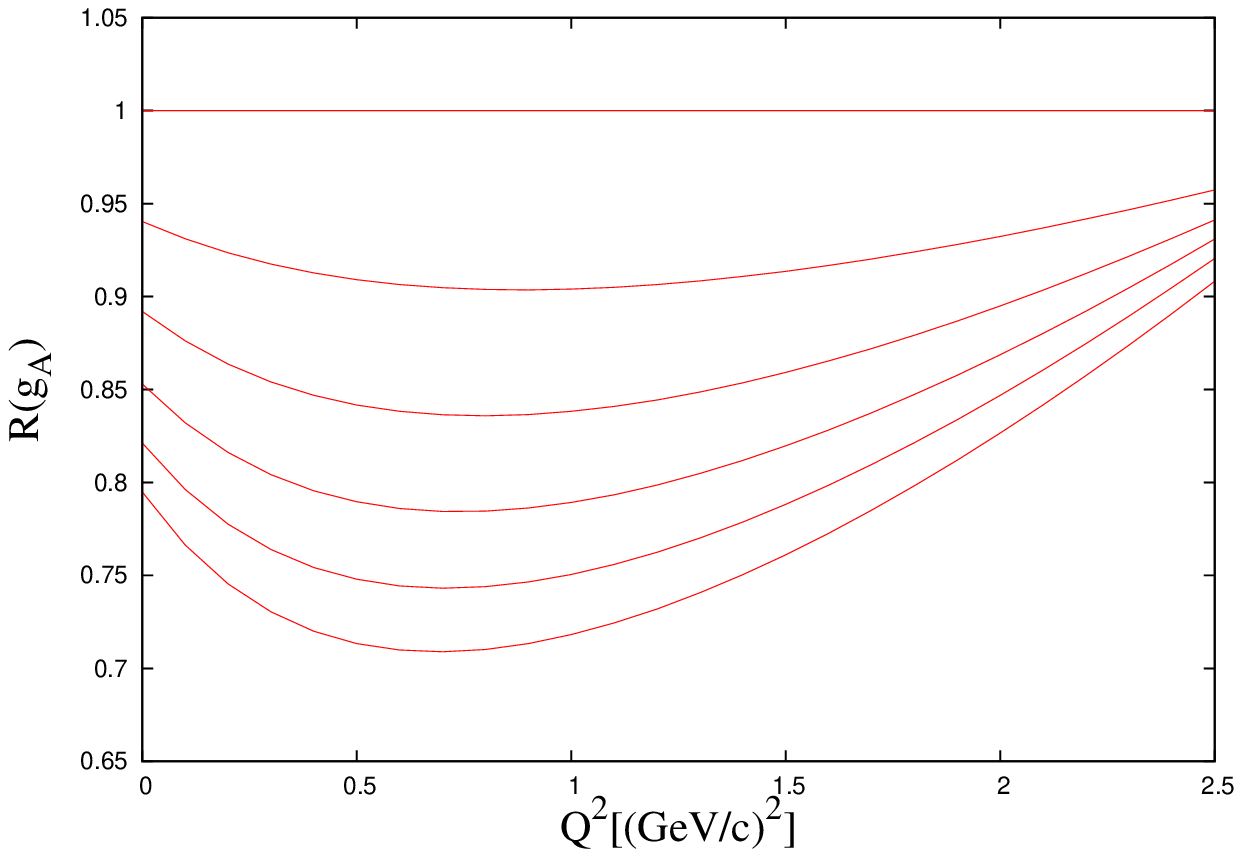}
\caption{(Color online) Effective nucleon mass $M^*(\rho)$ versus the nuclear density ratio $\rho / \rho_0$ (left),
and the in-medium axial coupling constant normalized to that in free space (right), $R(g_A) = g_A(\rho, Q^2)/g_A(\rho = 0, Q^2)$,
as a function of $Q^2$.  In the right panel, from the uppermost (for vacuum), the density increases
by $0.5 \rho_0$ in order. The lowermost curve is for $\rho = 2.5 \rho_0$.}~\label{fig1:massga}
\end{figure}
In Fig.~1 the effective nucleon mass is illustrated in the left panel, which shows a monotonic decrease of the mass with increasing
the nuclear density.
The modification of the axial coupling constant in nuclear matter is also shown in the right panel as a function of $Q^2$.
Even in the region of small momentum transfer, where most of the $\nu$ reactions expected
in the cosmos are assumed to occur, the reduction of the axial coupling constant $g_A (\rho, Q^2)$ (or equivalent to the axial form factor) amounts to
11\% at $\rho_0$.
Figure~2 presents the density dependence of the weak form factors given by Eq.~(2).
The modification of $F_2^{V, CC}$ at small $Q^2$ is more significant than that of $F_1^{V, CC}$, and, for example,
$F_2^{V, CC}$ is enhanced by about 11\% at $\rho_0$.
\begin{figure}
\centering
\includegraphics[width=8.0cm]{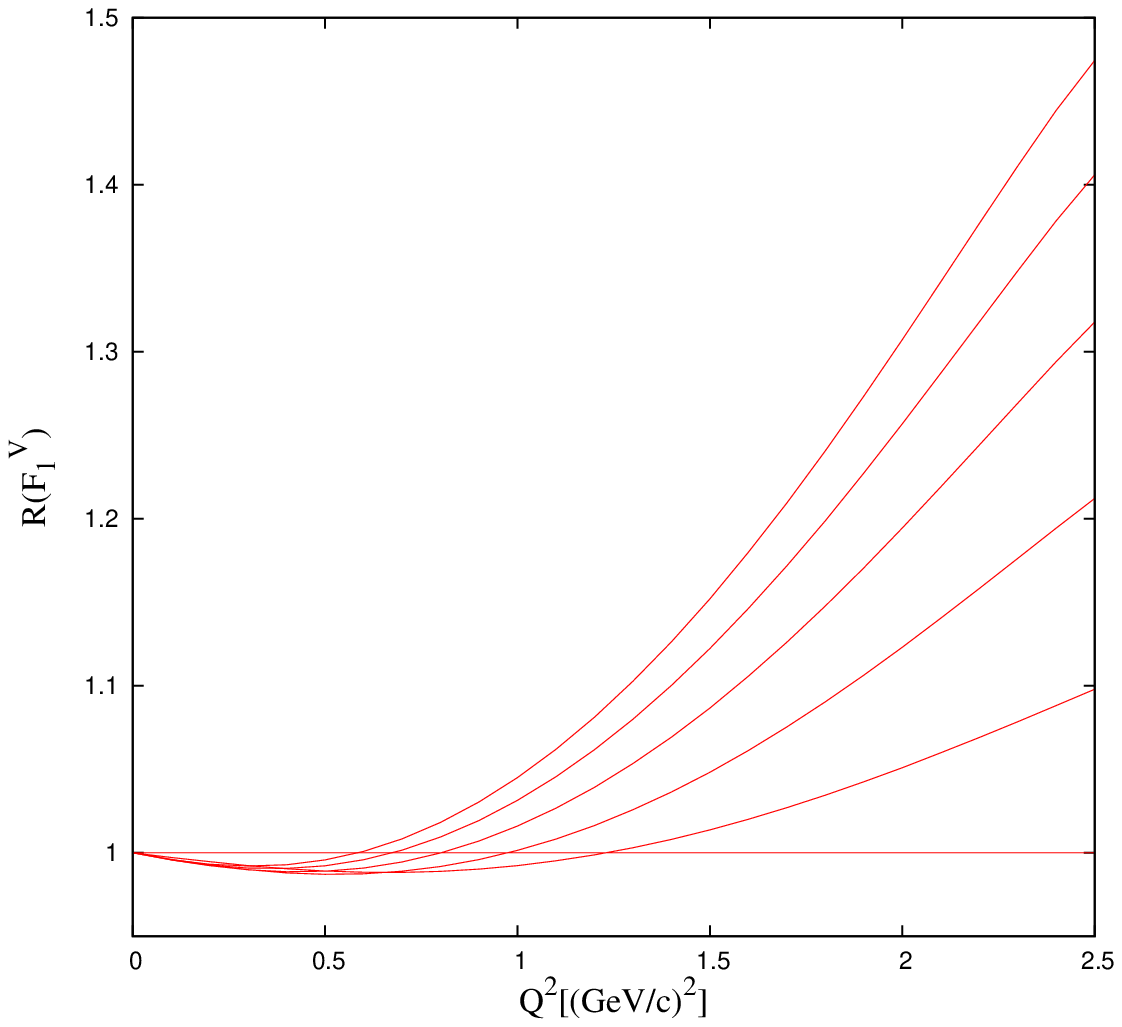}
\includegraphics[width=8.0cm]{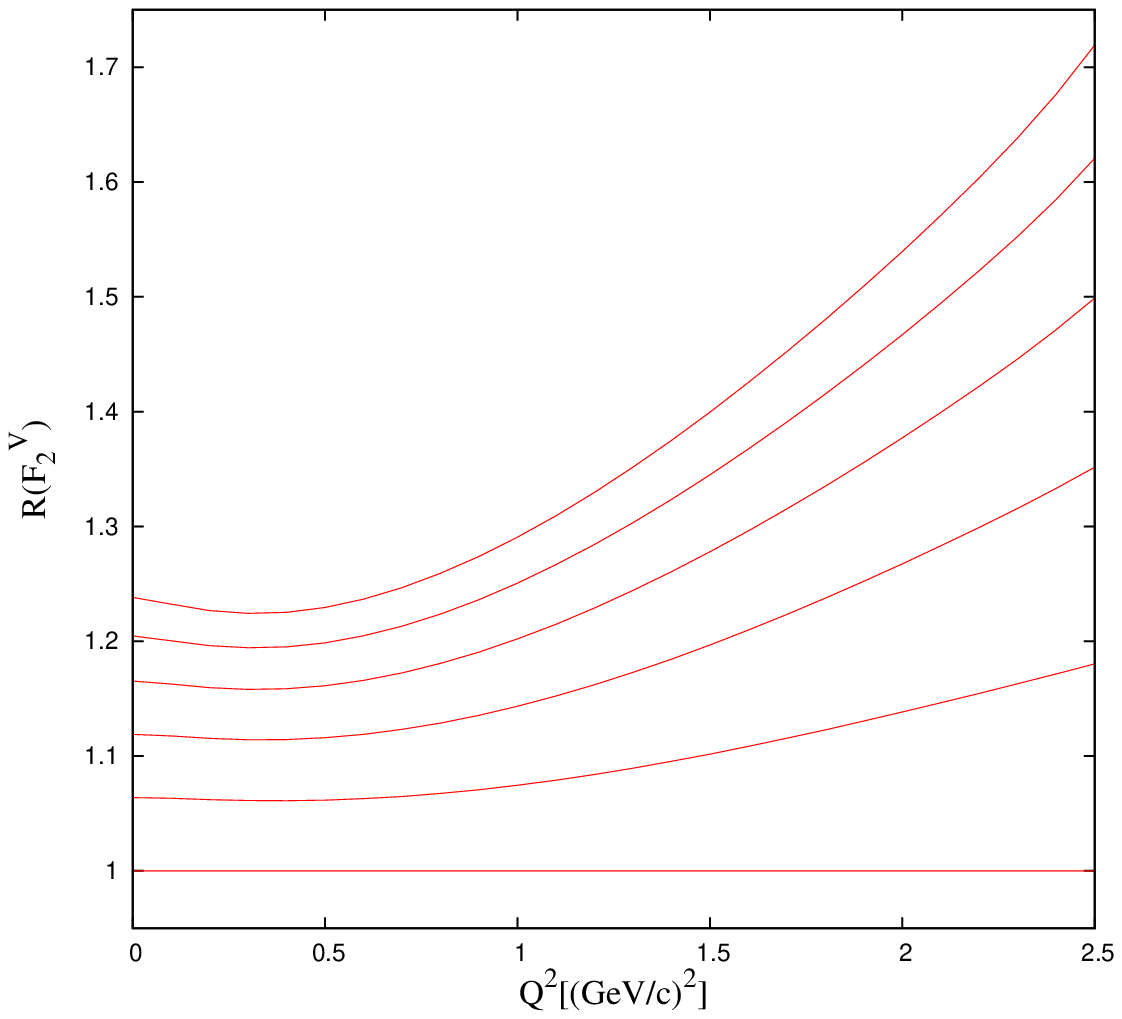}
\caption{Modification of the weak form factors, $ R(F_{1,2}^V) = F_{1,2}^{V, CC} (\rho, Q^2)/F_{1,2}^{V, CC}(\rho = 0, Q^2)$. From
the lowermost (for vacuum), the density increases by $0.5 \rho_0$ in order.
The uppermost curve is for $\rho = 2.5 \rho_0$. } \label{fig2:f1f2}
\end{figure}
%

Figure~3 shows the total cross sections for the ${\bar \nu_e} + p \rightarrow e^+ + n$ and ${ \nu_e} + n \rightarrow e^- + p $ reactions in nuclear matter.
Here, we have used the differential cross section formula with
the Sachs form factors, $G_M^{CC}$, $G_E^{CC}$ and $G_A^{CC} (= F_A^{CC})$~\cite{Ch08}:
\begin{eqnarray}
{( {  { d \sigma } \over {d Q^2  }} )}_{\nu ({\bar \nu})}^{CC} & = &
{ {G_F^2 ~ {cos}^2 \theta_c} \over { 2 \pi}} [ { 1 \over 2} y^2 {(G_M^{CC})}^2 + ( 1 - y - { y M \over {2 E} }) {   {{(G_E^{CC})}^2 + y E{(G_M^{CC})}^2/2M  }
\over { 1 + {y E \over { 2 M }} }}  \nonumber \\
& & + ( {y^2 \over 2} + 1 - y + { y M \over { 2 E}  }    ) {(G_A^{CC})}^2 \mp 2 y ( 1 - { y \over 2} ) G_M^{CC} G_A^{CC}],
\end{eqnarray}
where $y = Q^2 / (2 p \cdot k)$, $G_M^{CC} = F_1^{V,CC} + F_2^{V,CC}$, $G_E^{CC} = F_1^{V,CC} -{Q^2 \over { 4 M^2}} F_2^{V,CC}$, and
$E$ is the initial neutrino or anti-neutrino energy.
The signs $\mp$ respectively correspond to the ${\nu}_e \choose {\bar \nu}_e$ reactions, and the cross section is multiplied by
the Cabbibo-angle factor, $cos ^2 \theta_c$.  For both reactions, the present results in vacuum (black (solid) curves)
are consistent with the previous results calculated in Ref.~\cite{Stru03}.

For the ${\bar \nu_e} + p \rightarrow e^+ + n$ reaction in matter,
the total cross sections decrease by about 15\% per each increase (by $0.5 \rho_0$) of the density.
\begin{figure}
\centering
\includegraphics[width=8.0cm]{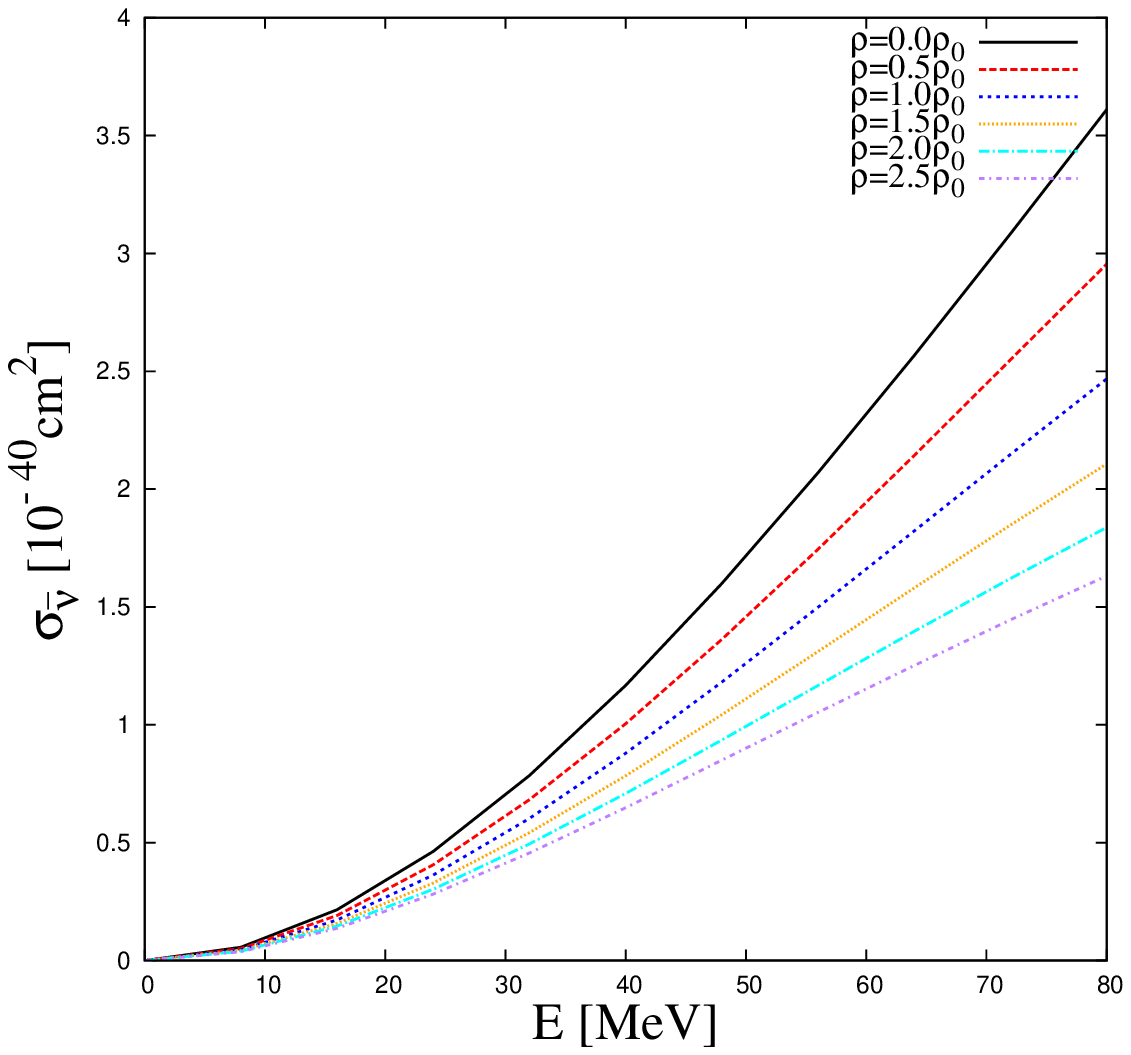}
\includegraphics[width=8.0cm]{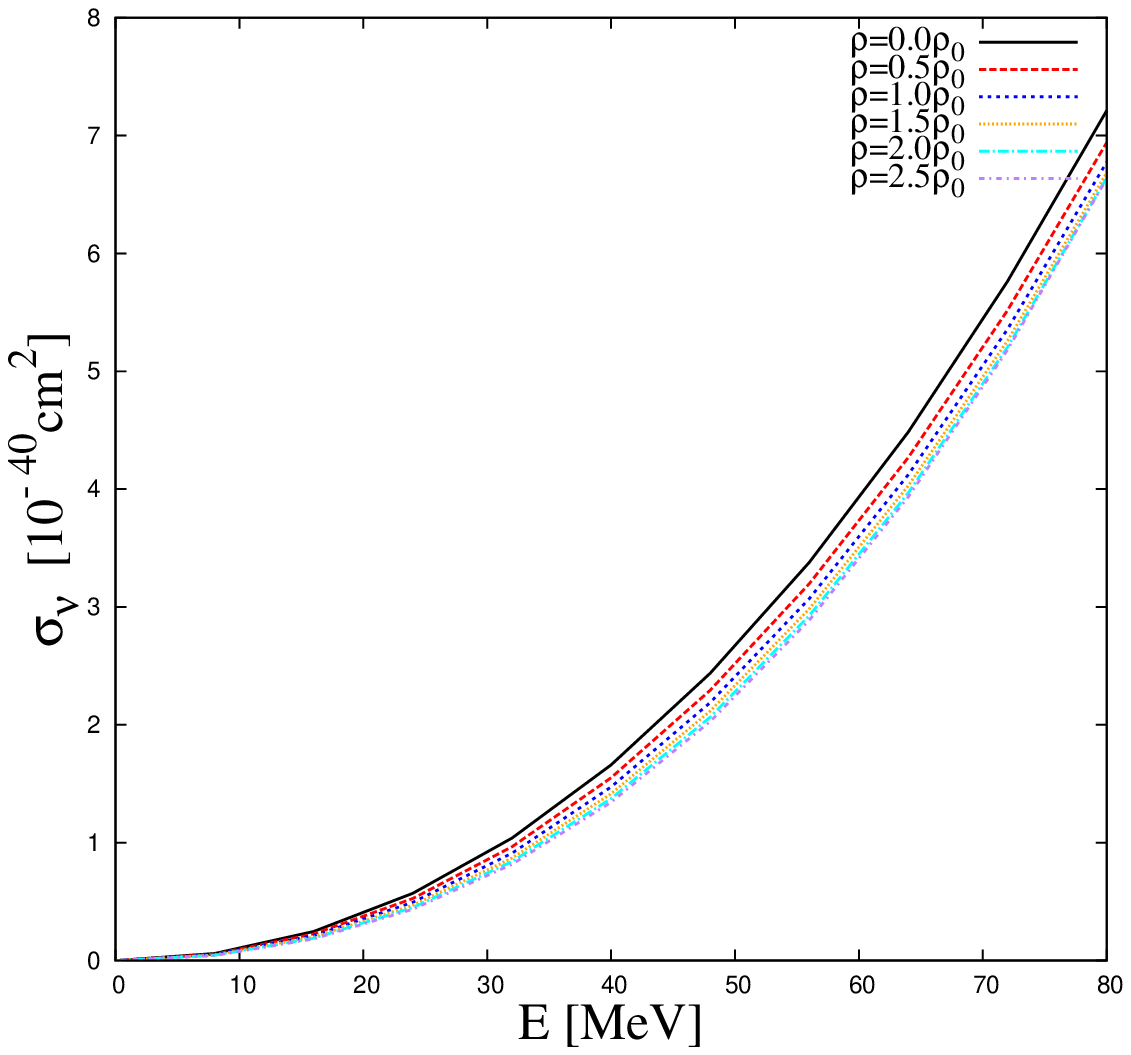}
\caption{(Color Online) Density dependence of the total cross sections for the
${\bar \nu_e} + p \rightarrow e^+ + n$ (left) and ${ \nu_e} + n \rightarrow e^- + p $ (right) in nuclear matter.
The black (solid) curves are for the results in free space. The cross sections in both reactions decrease with increasing the density:
$\rho / \rho_0$ = 0.5 (red (dashed)), 1.0 (blue (dotted)), 1.5 (yellow (short dotted)), 2.0 (cyan (dot-long-dashed)) and 2.5 (purple (dot-dashed)).} \label{fig3:nucleon}
\end{figure}
In contrast, for the ${ \nu_e} + n \rightarrow e^- + p $ reaction, the variation of the cross section in matter below $E_{\nu} \sim$ 30 MeV
is less than 2 \% (per each increase of the density).   Even around $E_{\nu} \sim$ 80 MeV, the cross section maximally decreases by about 5 \%
at $\rho_0$, compared to that in free space.

The large asymmetry between the ${\bar \nu}_e$ and ${\nu}_e$ reactions in nuclear matter can be explained
by the last term in Eq.~(3), namely the helicity-dependent (HD) term. 
In Fig.~4, we show the difference,
$\sigma^- = \sigma (\nu_e) - \sigma ( {\bar \nu}_e)$, and the sum, $\sigma^+ = \sigma (\nu_e) + \sigma ( {\bar \nu}_e)$, of the cross sections, which respectively
correspond to the HD and non-HD terms. 
The HD term in the left panel increases with increasing the density,
while the non-HD term in the right panel shows the decrease. 
For the ${\nu}_e$ reaction, which is given by a half of $\sigma^+ + \sigma^-$, the HD term plays a counterbalancing
role of the density effects, and it thus leads to the smaller density effect (see the right panel of Fig.~3).
In contrast, for the ${\bar \nu}_e$ reaction, the HD term enhances the density effect on the cross section.
Therefore, the large difference in the density effects on the $\nu$ and ${\bar \nu}$ reactions is closely associated with the difference
in the helicities of the incident $\nu$ and ${\bar \nu}$.

The change of the form factors definitely affects the ${\bar \nu}_e -$ and ${\nu}_e -$nucleon interactions in dense nuclear matter.  As a result,
around $\rho_0$, the variation of the ${\bar \nu}_e ({\nu}_e)-$nucleon cross section maximally amounts to 35 (5) \%.
The radiative corrections and the Coulomb distortion 
are not taken into account in this work, because those effects are known to be less than 2 \%~\cite{Stru03}.

\begin{figure}
\centering
\includegraphics[width=8.0cm]{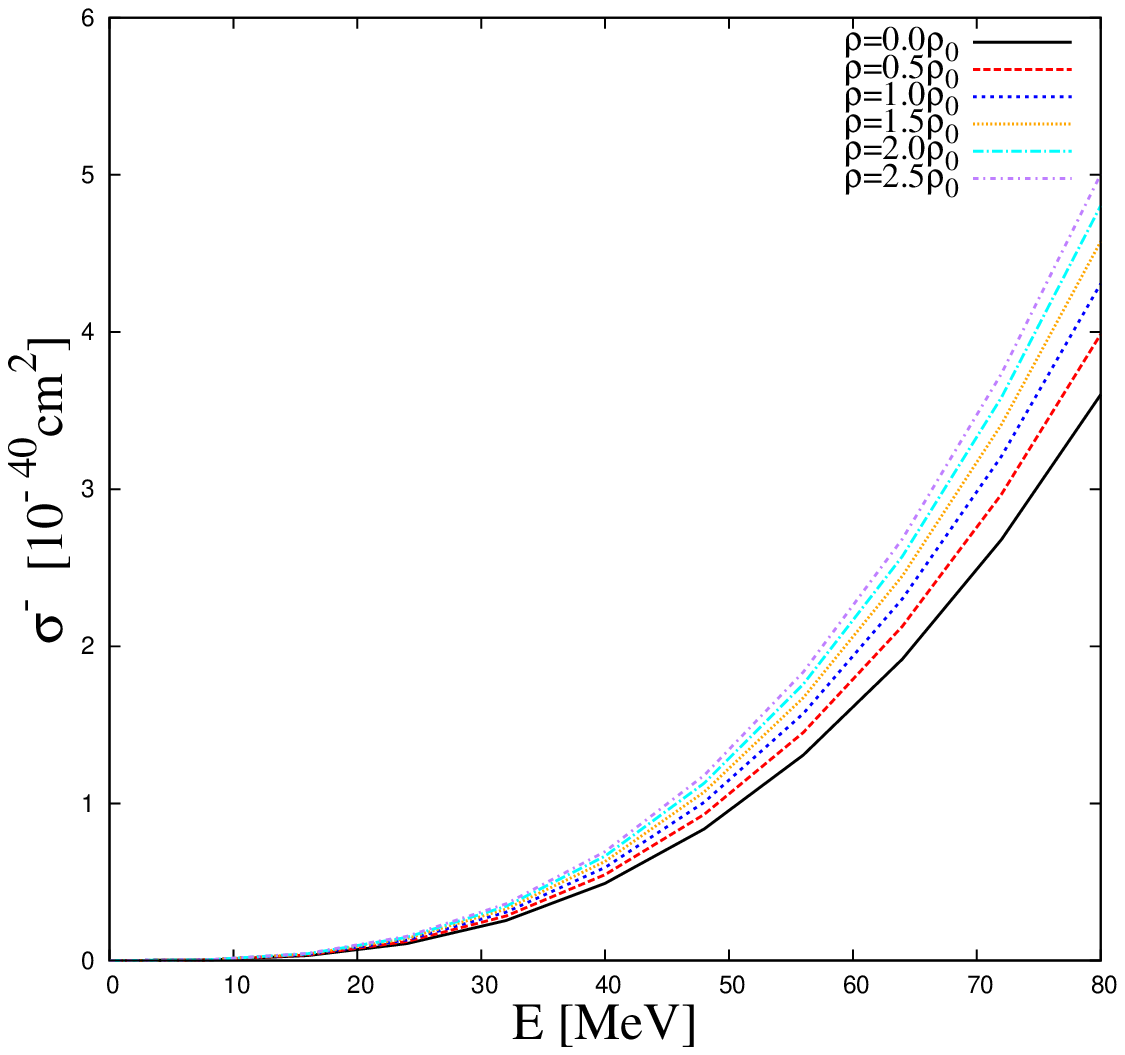}
\includegraphics[width=8.0cm]{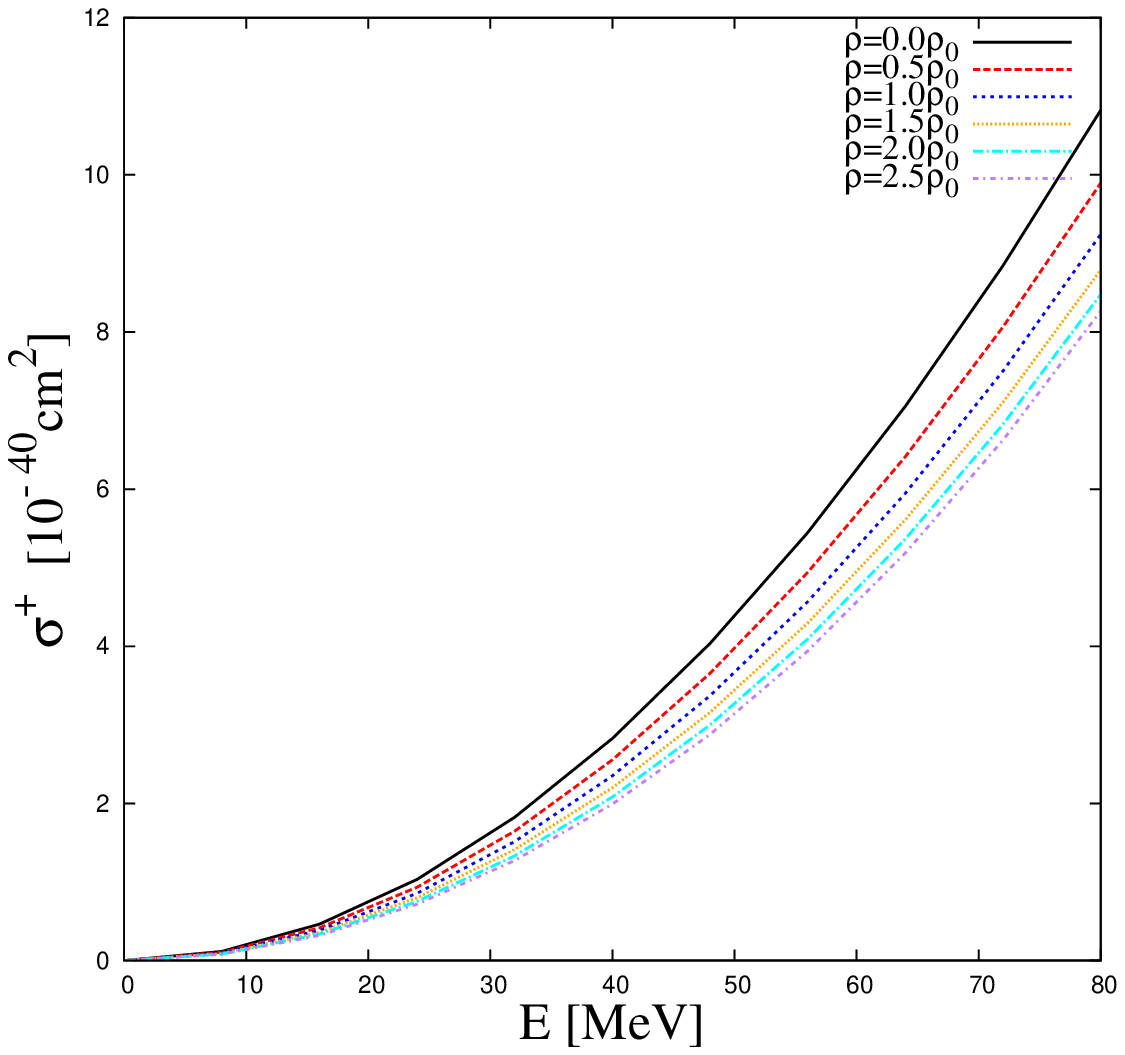}
\caption{(Color Online) Difference (left), $\sigma^- =\sigma (\nu_e) - \sigma ( {\bar \nu}_e)$, and
the sum (right), $\sigma^+ =\sigma (\nu_e) + \sigma ( {\bar \nu}_e)$, of the total cross sections.
}
\label{fig4:asymnucleon}
\end{figure}
%

%
%
%

In order to investigate the asymmetry in the $\nu_e$ and ${\bar \nu}_e$ reactions on $^{12}$C, we use the following differential cross
section formula, which is explained in detail in Ref.~\cite{ch10},
\begin{eqnarray} & & ({{d \sigma_{\nu}} \over {d \Omega }  })_{(\nu / {\bar
\nu})} = { { G_F^2 \epsilon k ~ {cos}^2 \theta_c } \over {\pi ~ (2 J_i + 1 ) }}~
\bigl[ ~ {\mathop\Sigma_{J = 0}} (
 1+ {\vec \nu} \cdot {\vec \beta }){| <  J_f || {\hat {\cal M}}_J || J_i > | }^2
 \\ \nonumber & & + (
 1 - {\vec \nu} \cdot {\vec \beta } + 2({\hat \nu} \cdot {\hat q} )
 ({\hat q} \cdot {\vec \beta}  ))
  {| <  J_f || {\hat {\cal L}}_J ||
J_i > | }^2  \\ \nonumber
& & - {\hat q} \cdot ({\hat \nu}+ {\vec \beta} )  { 2 Re < J_f || {\hat {\cal L}}_J  || J_i>
{< J_f|| {\hat {\cal M}}_J || J_i >}^*  } \\
\nonumber & &  + {\mathop\Sigma_{J = 1}} ( 1 - ({\hat \nu} \cdot
{\hat q} )({\hat q} \cdot {\vec \beta}  ) ) ( {| <  J_f || {\hat
{\cal T}}_J^{el}  || J_i > | }^2 + {| <  J_f || { {\hat
{\cal T}}}_J^{mag} || J_i > | }^2
) \\
\nonumber & &  \pm {\mathop\Sigma_{J = 1}} {\hat q} \cdot ({\hat
\nu} - {\vec \beta} )  2 Re [ <  J_f || { {\hat {\cal T}}}_J^{mag} ||
J_i > {<  J_f || {{\hat {\cal T}}}_J^{el} || J_i > }^* ]\bigr]~,
\end{eqnarray}
where ${\vec \nu}$ and $ {\vec k}$ are respectively the 3-momenta of the initial and final
leptons, ${\vec q} = {\vec k} - {\vec \nu}$, ${\vec \beta} =
{\vec k} / \epsilon $ with $\epsilon$ being the energy of the final lepton, and the density-dependent form factors are involved in the
reduced matrix elements.  We here include the Coulomb
distortion on the outgoing leptons due to the residual nucleus~\cite{ch10-2}.
\begin{figure}
\centering
\includegraphics[width=8.0cm]{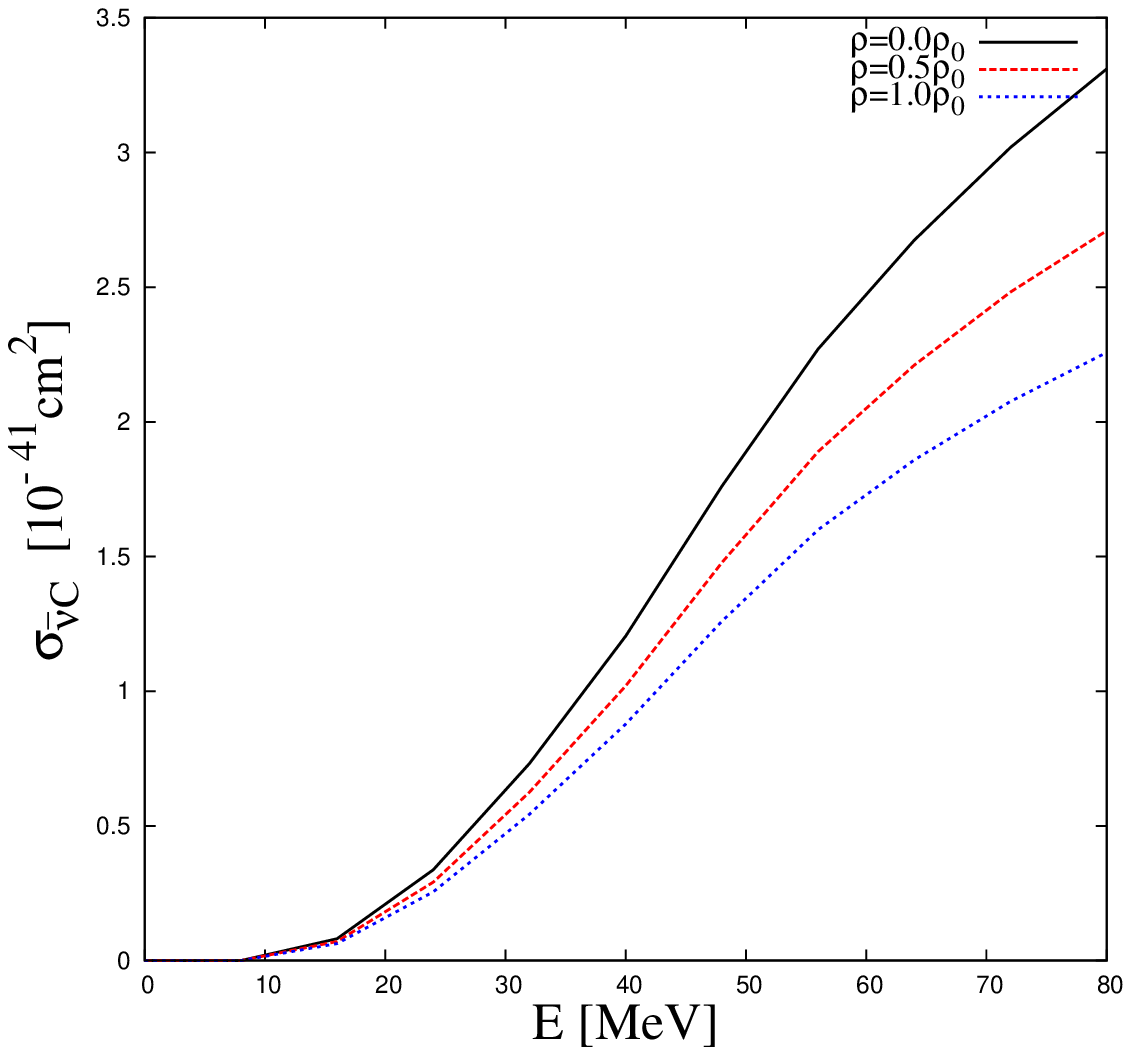}
\includegraphics[width=8.0cm]{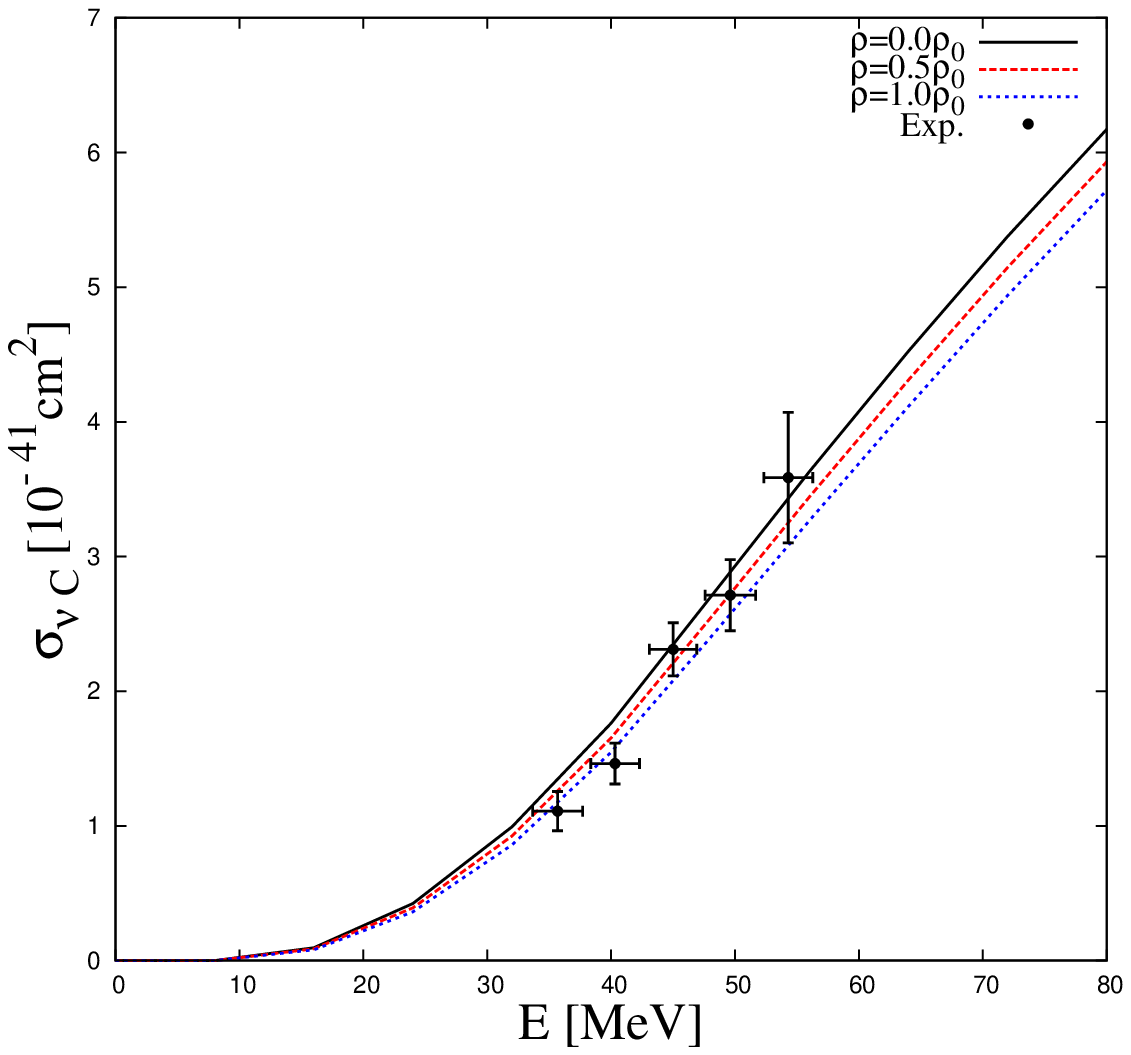}
\caption{(Color Online) Density dependence of the total cross sections for $^{12}$C$({\bar \nu}_e, e^+ ) ^{12}$B$_{g.s. (1^+)}$ (left)
and $^{12}$C$(\nu_e, e^- ) ^{12}$N$_{g.s. (1^+)}$ (right). The result denoted by the black (solid) curve (for vacuum) in the right panel is
the same as that of the $1^+$ transition shown in Fig.~1 of Ref.~\cite{ch10-2}.
The measured cross section~\cite{Athan97} is also presented in the right panel.
} \label{fig5:c12}
\end{figure}

Using Eq.~(4), we calculate the total cross sections for $^{12}$C$({\bar \nu}_e, e^+ ) ^{12}$B$_{g.s. (1^+)}$ and  $^{12}$C$(\nu_e, e^- )^{12}$N$_{g.s. (1^+)}$.
Within the QRPA framework~\cite{ch10-2}, the reactions can be treated by the $\Delta J = 1$ transition from the $0^+$ ground state of $^{12}$C to the $1^+$ ground
states of $^{12}$B or $^{12}$N.  The numerical results are then presented in Fig.~5.  To see the effect of the density-dependent form factors on the cross section,  we assume that
the nuclear density in the interior of $^{12}$C is constant ($\rho/\rho_0 = 0, 0.5, 1.0$), and calculate the form factors in Eq.~(4) at those densities.

The large asymmetry in the ${\bar \nu}_e -$ and ${\nu}_e - ^{12}$C cross sections due to the medium effect appears
in a similar fashion to that in the case of the in-medium nucleon shown in Fig.~3.
We note that, because we consider only the exclusive reaction to the ground state of the daughter nuclei,
the cross sections are smaller than those for the nucleon in Fig.~3.

\begin{figure}
\centering
\includegraphics[width=8.0cm]{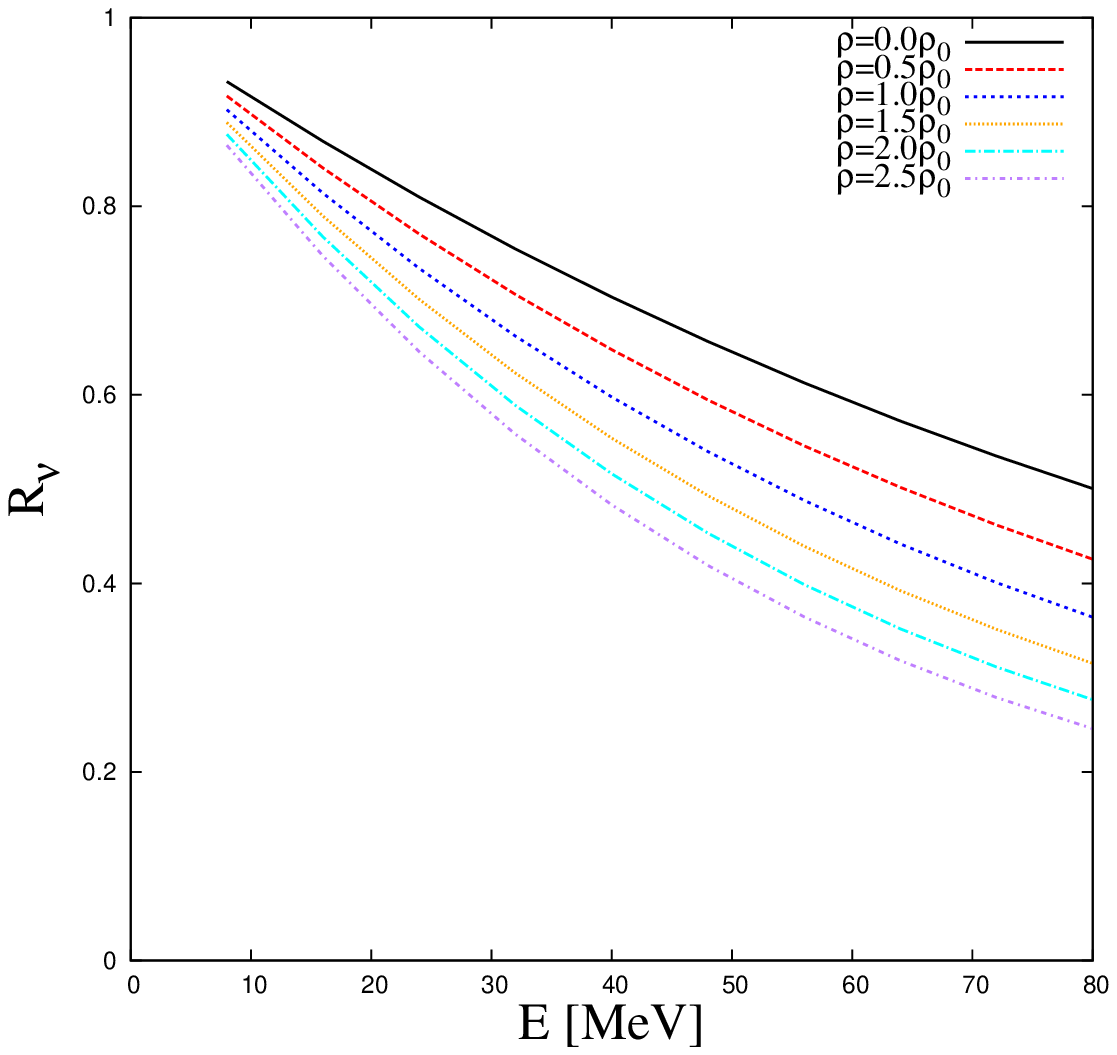}
\includegraphics[width=8.0cm]{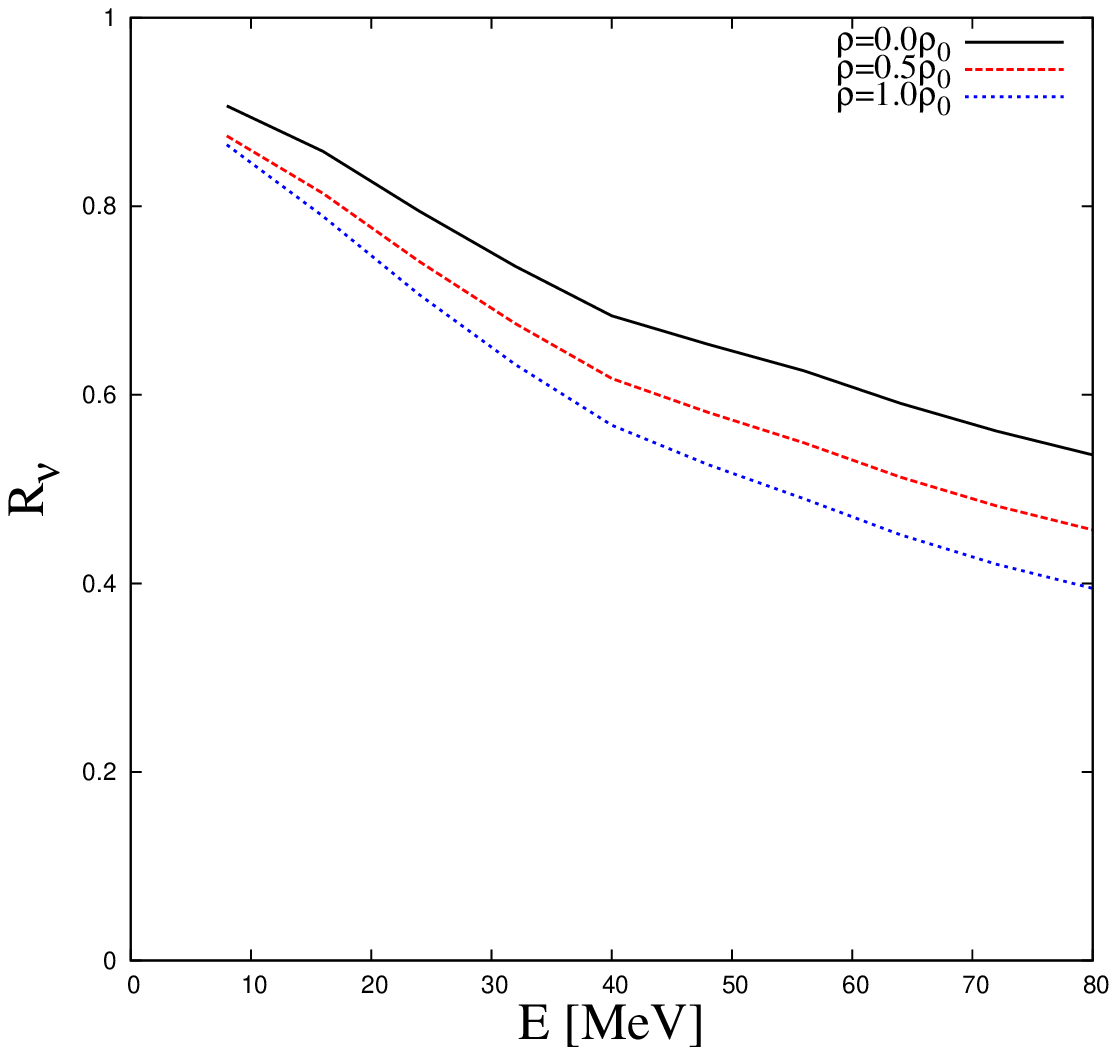}
\caption{(Color Online) Energy dependence of the ratio of the ${\bar \nu}_e$ cross section to the $\nu_e$ one, $R_{\nu} = \sigma ({\bar \nu}_e) /  \sigma ({\nu}_e)$.
The left panel is for a nucleon in nuclear matter, while the right one is for $^{12}$C.
} \label{fig6:ratio}
\end{figure}

The energy dependence of the measured cross section for $^{12}$C$(\nu_e, e^-) ^{12}$N$_{g.s. (1^+)}$~\cite{Athan97} is also shown in the right panel of Fig.~5.
Contrary to the significant reduction in the ${\bar \nu}_e - ^{12}$C cross section, the
$\nu_e - ^{12}$C cross section around the saturation density $\rho_0$ is reduced less than 10 \%,
comparing with that in free space. Furthermore, if we take the average Fermi momentum of
$^{12}$C, which is $k_F = 225$ MeV ($\rho = 0.668 \rho_0$)~\cite{Kim04}, to calculate the form factors,
the maximum change due to the bound nucleon is shown to be less than 5\%,
which is within the experimental error bars. Since the Coulomb distortion
effect is known to be of about 5 $\sim$ 6 \%~\cite{ch10-2},
the density effect on the $\nu_e - ^{12}$C reaction is as large as the Coulomb distortion.

Finally, in Fig.~6, we show the initial-energy dependence of the cross-section ratio, $R_{\nu} = \sigma ({\bar \nu}_e) /  \sigma ({\nu}_e)$, on a nucleon
in nuclear matter or in $^{12}$C.  The ratio decreases with increasing the energy $E$.
This can be regarded as a valuable, low-energy extension of the previous calculation performed by Kuzmin et al.~\cite{Kuzmin06},
in which $E$ is assumed to be larger than 100 MeV.  As for the density effect, the higher the density increases,
the smaller the ratio becomes.  It means that the asymmetry between the ${\bar \nu}_e$ and ${\nu}_e$ reactions is expected to become larger
in a denser nuclear medium, irrespective of the target.


As a summary, we have studied the $\nu_e$ and ${\bar \nu}_e$ induced reactions via charged current
on a nucleon in nuclear matter or in $^{12}$C, using the density-dependent weak form factors
calculated by the QMC model. 
The present calculation has revealed that the anti-neutrino cross section is reduced maximally by about 35 \% at the saturation density $\rho_0$, while
the neutrino cross section at the same density
is reduced by about 5 \% at most. Therefore, it is expected that this asymmetry in the $\nu_e$ and ${\bar \nu}_e$ induced reactions influences
the r-process and also the neutrino-process nucleosynthesis in core-collapse supernovae.
In particular, the anti-neutrino propagation inside the PNS is significantly affected by the density effect caused by
the variation of the nucleon structure in dense matter.  The $\nu -$ and ${\bar \nu} - ^{12}$C cross
sections also show large asymmetry, namely the anti-neutrino cross section is reduced much, while the reduction of the neutrino cross section is rather
small. For instance, the cross section for $^{12}$C$(\nu_e, e^- ) ^{12}$N$_{g.s. (1^+)}$ is reduced by about 5 \% around $\rho_0$,
which is as large as the amount of the Coulomb distortion effect.




\begin{acknowledgments}
This work was supported by the National Research Foundation of Korea (Grant No. 2011-0015467).
The work of KT was supported by the University of Adelaide and the Australian Research Council through
grant No.~FL0992247 (AWT).
\end{acknowledgments}



\thebibliography{99}

\bibitem{Fisc11} T. Fischer, I. Sagert, G. Pagliara, M. Hempel, J. Schaffner-Bielich, T. Rauscher, F.-K. Thielemann, R. Kaeppeli, G. Martinez-Pinedo,
and M. Liebendoerfer, Astrophys. J. Supplement {194} (2011) 39.
\bibitem{Maru11} Tomoyuki Maruyama, Toshitaka Kajino, Nobutoshi Yasutake, Myung-Ki Cheoun, Chung-Yeol Ryu, Phys. Rev. { D 83}  (2011) 081303(R).
\bibitem{Maru12} Tomoyuki Maruyama, Nobutoshi Yasutake, Myung-Ki Cheoun, Jun Hidaka, Toshitaka Kajino, Grant J. Mathews, Chung-Yeol Ryu, Phys. Rev. { D 86} (2012)   123003.
\bibitem{Wana11} Shinya Wanajo, Hans-Thomas Janka, Shigeru Kubono,
Astrophys. J. { 729}  (2011) 46.



\bibitem{Ch10-2} Myung-Ki Cheoun, Eunja Ha, T. Hayakawa, Toshitaka Kajino, Satoshi Chiba, Phys. Rev. { C82} (2010) 035504 .

\bibitem{Ch12} Myung-Ki Cheoun, Eunja Ha, T. Hayakawa, Satoshi Chiba, Ko Nakamura, Toshitaka Kajino, Grant J. Mathews, Phys. Rev. { C85} (2012) 065807.

\bibitem{Yosh08} T. Yoshida, T. Suzuki, S, Chiba, T. Kajino, H. Yokomukura,
K. Kimura, A. Takamura, H. Hartmann, Astrophys. J. { 686} (2008) 448.

\bibitem{Malo00} S. Malov, K. Wijesooriya, F. T. Baker, L. Bimbot,
E. J. Brash, C. C. Chang, J. M. Finn, K. G. Fissum
et al., Phys. Rev. { C 62} (2000) 057302.
\bibitem{Diet01} S. Dieterich, P. Bartsch, D. Baumann, J. Bermuth,
K. Bohinc, R. Bohm, D. Bosnar, S. Derber et al., Phys.
Lett. { B 500} (2001) 47.
\bibitem{Stra03} S. Strauch et al. [ Jefferson Lab E93-049 Collaboration ],
Phys. Rev. Lett. { 91} (2003) 052301.
\bibitem{Paol10} M. Paolone, S. P. Malace, S. Strauch, I. Albayrak, J. Arrington,
B. L. Berman, E. J. Brash, B. Briscoe et al.,
Phys. Rev. Lett. { 105} (2010) 072001.
\bibitem{Mala11} S. P. Malace, M. Paolone, S. Strauch, I. Albayrak, J. Arrington,
B. L. Berman, E. J. Brash, B. Briscoe et al.,
Phys. Rev. Lett. { 106} (2011) 052501.
\bibitem{Cloe09} I. C. Cloet, G. A. Miller, E. Piasetzky and G. Ron, Phys.
Rev. Lett. { 103} (2009) 082301.



\bibitem{QMCboundff}
  D.~-H.~Lu, K.~Tsushima, A.~W.~Thomas, A.~G.~Williams and K.~Saito,
  Phys.\ Rev.\ C { 60} (1999) 068201.

\bibitem{QMChe3ff}
  D.~-H.~Lu, K.~Tsushima, A.~W.~Thomas, A.~G.~Williams and K.~Saito,
  Phys.\ Lett.\ B { 441} (1998) 27.

\bibitem{QMCmatterff}
  D.~-H.~Lu, A.~W.~Thomas, K.~Tsushima, A.~G.~Williams and K.~Saito,
  Phys.\ Lett.\ B { 417} (1998) 217.

\bibitem{QMCmatterga}
  D.~-H.~Lu, A.~W.~Thomas and K.~Tsushima,
  nucl-th/0112001 (2011).


\bibitem{QMCfinite}
K.~Saito, K.~Tsushima and A.~W.~Thomas,
  Nucl.\ Phys.\ A { 609} (1996) 339;\\
K.~Saito, K.~Tsushima and A.~W.~Thomas,
  Phys.\ Rev.\ C { 55} (1997) 2637.


\bibitem{QMChyp}
K.~Tsushima, K.~Saito, J.~Haidenbauer and A.~W.~Thomas,
  Nucl.\ Phys.\ A { 630} (1998) 691;\\
P.~A.~M.~Guichon, A.~W.~Thomas and K.~Tsushima,
  Nucl.\ Phys.\ A { 814} (2008) 66;\\
K.~Tsushima and F.~C.~Khanna,
  Phys.\ Rev.\ C { 67} (2003) 015211;\\
K.~Tsushima and F.~C.~Khanna,
  J.\ Phys.\ G { 30} (2004) 1765.


\bibitem{QMCreview}
  K.~Saito, K.~Tsushima and A.~W.~Thomas,
  Prog.\ Part.\ Nucl.\ Phys.\  { 58} (2007) 1.  

\bibitem{Kim04} K. Tsushima, Hungchong Kim, and K. Saito, Phys. Rev. { C 70} (2004) 038501.





\bibitem{ch10} Myung-Ki Cheoun, Eunja Ha, K.S. Kim, Toshitaka Kajino, J. Phys. { G 37} (2010) 055101.
\bibitem{ch10-2} Myung-Ki Cheoun, Eunja Ha, Su Youn Lee, K.S. Kim, W.Y. So, Toshitaka Kajino, Phys. Rev. { C81} (2010) 028501.

\bibitem{giusti1} Andrea Meucci, Carlotta Giusti, and Franco Davide
Pacati, Nucl. Phys. { A739} (2004) 277; Nucl. Phys. {A744}, 307 (2004); Nucl. Phys. { A773} (2006) 250.
\bibitem{musolf}M. J. Musolf and T. W. Donnelly, Nucl. Phys. {
A546} (1992) 509.


\bibitem{Ch08} Myung-Ki Cheoun and K.S. Kim, J. Phys. { G 35} (2008) 065107.

\bibitem{Stru03} Alessandro Strumia and Francesco Vissani, Phys. Lett. { B 564} (2003) 42.


\bibitem{Athan97} C. Athanassopoulos et al. (LSND Collaboration), Phys. Rev. { C 55} (1997) 2078.

\bibitem{Kuzmin06} K. S. Kuzmin, V. V. Lyubushkin, and V. A. Naumov, Physics of Atomic Nuclei { 69} (2006) 1857.
\end{document}